\documentclass[10pt,conference]{IEEEtran}
\IEEEoverridecommandlockouts
\pdfoutput=1
\usepackage[T1]{fontenc}
\usepackage{cite}
\usepackage{amsmath,amssymb,amsfonts}
\usepackage{algorithm}
\usepackage[noend]{algpseudocode}
\usepackage{caption}
\usepackage{graphicx}
\usepackage{textcomp}
\usepackage{array}
\usepackage{multirow}
\usepackage{multicol}
\usepackage{tabulary}
\usepackage{booktabs}
\usepackage{dcolumn} 
\usepackage{bm} 
\usepackage{braket}
\usepackage[version=4]{mhchem}
\usepackage{hyperref} 
\usepackage{siunitx} 
\usepackage[normalem]{ulem} 
\hypersetup{
    colorlinks = true,
    citecolor = magenta,
    linkcolor = purple
}
\def\BibTeX{{\rm B\kern-.05em{\sc i\kern-.025em b}\kern-.08em
    T\kern-.1667em\lower.7ex\hbox{E}\kern-.125emX}}
    
\usepackage[normalem]{ulem}

\usepackage[dvipsnames, table]{xcolor}
\usepackage{tabularx, makecell, linegoal}

\usepackage{ifthen}

\newboolean{ShowComments}
\setboolean{ShowComments}{true}  
\ifthenelse{\boolean{ShowComments}}%
	{
		\newcommand{\ColorComment}[3]{%
				{\colorbox{#1}{\color{White}   \textsf{\textbf{#2}}} \textcolor{#1}{#3}}}

	}%
	{
		\newcommand{\ColorComment}[3]{}

	}%

\definecolor{rdvcolor}{rgb}{0,0.5,0}
\definecolor{satohcolor}{RGB}{254,0,0}
\definecolor{michalcolor}{RGB}{255,127,80}
\definecolor{naphanncolor}{RGB}{112, 51, 173}

\begin{document}
\bstctlcite{IEEEexample:BSTcontrol}


\title{Engineering Challenges in All-photonic Quantum Repeaters}

\author{
\IEEEauthorblockN{Naphan Benchasattabuse\IEEEauthorrefmark{1}\IEEEauthorrefmark{4},
Michal Hajdu\v{s}ek\IEEEauthorrefmark{1}\IEEEauthorrefmark{4},
and Rodney Van Meter\IEEEauthorrefmark{3}\IEEEauthorrefmark{4}}\\ 
\IEEEauthorblockA{\IEEEauthorrefmark{1}\textit{Graduate School of Media and Governance, Keio University Shonan Fujisawa Campus, Kanagawa, Japan}}
\IEEEauthorblockA{\IEEEauthorrefmark{3}\textit{Faculty of Environment and Information Studies, Keio University Shonan Fujisawa Campus, Kanagawa, Japan}}
\IEEEauthorblockA{\IEEEauthorrefmark{4}\textit{Quantum Computing Center, Keio University, Kanagawa, Japan}\\
\{whit3z,michal,rdv\}@sfc.wide.ad.jp}
}

\thispagestyle{plain}
\pagestyle{plain}
\maketitle

\begin{abstract}
Quantum networking, heralded as the next frontier in communication networks, envisions a realm where quantum computers and devices collaborate to unlock capabilities beyond what is possible with the Internet.
A critical component for realizing a long-distance quantum network, and ultimately, the Quantum Internet, is the quantum repeater. 
As with the race to build a scalable quantum computer with different technologies, various schemes exist for building quantum repeaters.
This article offers a gentle introduction to the two-way ``all-photonic quantum repeaters,'' a recent addition to quantum repeater technologies.
In contrast to conventional approaches, these repeaters eliminate the need for quantum memories, offering the dual benefits of higher repetition rates and intrinsic tolerance to both quantum operational errors and photon losses.
Using visualization and simple rules for manipulating graph states, we describe how all-photonic quantum repeaters work.
We discuss the problem of the increased volume of classical communication required by this scheme, which places a huge processing requirement on the end nodes.
We address this problem by presenting a solution that decreases the amount of classical communication by three orders of magnitude.
We conclude by highlighting other key open challenges in translating the theoretical all-photonic framework into real-world implementation, providing insights into the practical considerations and future research directions of all-photonic quantum repeater technology.
\end{abstract}

\section{Introduction}

The Internet has transformed how we interact with each other and with the world to the point that it has become an indispensable part of our lives.
The possibility of connecting quantum devices together in a similar fashion has sparked the interest of scientists, who imagine the impact and the new capabilities that the Quantum Internet will unlock~\cite{wehner-vision-road-ahead, rfc9340, rdv-qi-architecture}.
Long-range entanglement, the primary resource in quantum networks, can enable a multitude of industrially and scientifically useful applications.
These include generating secret keys to make communications over the Internet more secure, making better sensors such as improving the resolution of telescopes, increasing the computational power of quantum computers by connecting them together, or executing blind quantum computations where the programs are sent to be computed off-site but nothing about the program, the input, or the output are known to the off-site server.

The quantum network, akin to the classical computer network, connects quantum computers or quantum devices together.
However, the basic task of a quantum network is not merely sending or receiving quantum data; it is distributing generic entangled states between two or more distant parties in the network.
These generic entangled states can then be consumed to execute the applications previously mentioned, including transferring data.
Sharing these generic entangled states equates to establishing quantum communication channels.
Each entangled state is single-use, thus many applications consume a large quantity of this resource.
Therefore, distributing high-quality entanglement at a fast rate is crucial in realizing a usable quantum network.

The smallest entangled state that a quantum network needs to be able to distribute, which can be used to build larger entangled states, is the Bell pair\footnote{A Bell pair is a maximally entangled bipartite state, a state where measuring one qubit gives a complete description of the state of the other qubit.}.
Although it would be more efficient for the network to generate generic multipartite entangled states, it makes managing the network layer itself much more difficult. 
We consider a network that only distributes Bell pairs, and if the application requires larger entangled states, the application layer can create them from these Bell pairs. 

The primary mechanism for generating long-distance Bell pairs between network nodes relies on exchanging single photons.
Attenuation of optical signals in fiber leads to an exponentially vanishing probability of photon arrival as the distance between network nodes increases, limiting practical quantum communication via direct photon transmission to only a few tens of kilometers.
To overcome this distance and scaling limit, quantum repeaters were introduced~\cite{briegel-repeater}.
Unlike the classical repeaters, where signals are amplified or regenerated, arbitrary quantum states cannot be copied due to the ``no-cloning'' theorem. 
Instead, with the help of stationary quantum memories\footnote{We use the term stationary quantum memory to emphasize that the qubits are located in the repeaters.}, quantum repeater nodes store the link-level Bell pairs -- Bell pairs generated via optical fiber -- and join them together to make them span multiple hops.

It is natural to think that quantum networks and quantum repeaters cannot be built before a working quantum computer because both share a need for controllable, long-coherence-time quantum memories.
This was shown to be in fact not necessary in the last decade by Azuma \emph{et al.}~\cite{azuma-rgs}, who introduced an all-photonic scheme without any quantum memories.
This seminal result led to a new variation of quantum repeaters~\cite{hasegawa-passive-absa-experiment,jian-wei-pan-rgs-experiment} whose development can be independent of the development of quantum computers or quantum memories.
Although this transition shifts the hardware challenge to the creation of a deterministic and controllable source of indistinguishable photons, it concurrently broadens the spectrum of methodologies for realizing quantum repeaters.

It is crucial to note that a quantum network is a hybrid system and that classical networking is an indispensable part of the quantum communication infrastructure.
Classical channels are used to transmit information about measurement outcomes, enabling error correction, and establishing entanglement between distant quantum nodes.
Therefore, while quantum repeaters represent a significant stride toward scalable quantum communication networks, their seamless integration with classical networking components remains a fundamental requirement for the realization of practical and reliable long-distance quantum communication.

In particular, classical communication plays a major role in the all-photonic repeater scheme, potentially imposing limitations on the overall Bell pair generation rate.
The scheme introduces redundancy to photonic qubits through quantum error-correcting codes to forego the necessity for quantum memories.
However, the ratio of the photons to end-to-end Bell pairs can span several orders of magnitude.
All the generated photons must be measured and tracked in order to create a Bell pair between two end nodes, thus necessitating a high bandwidth and processing power at the end nodes.

The contributions of this article are summarized as follows.
We first give a tutorial on the all-photonic quantum repeater scheme based on the \emph{repeater graph state} (RGS), proposed by Azuma \emph{et al.}~\cite{azuma-rgs}.
We continue by highlighting a particular disadvantage of this scheme, namely that better performance requires ever larger classical information that must be communicated and processed by the end nodes.
We propose a communication protocol that reduces the bandwidth and the data that end nodes are required to process.
We end the article by identifying some of the key open problems and potential solution ideas in realizing the RGS scheme.

\section{Quantum Repeaters}

\begin{figure*}
    \centering
    \includegraphics[width=\textwidth]{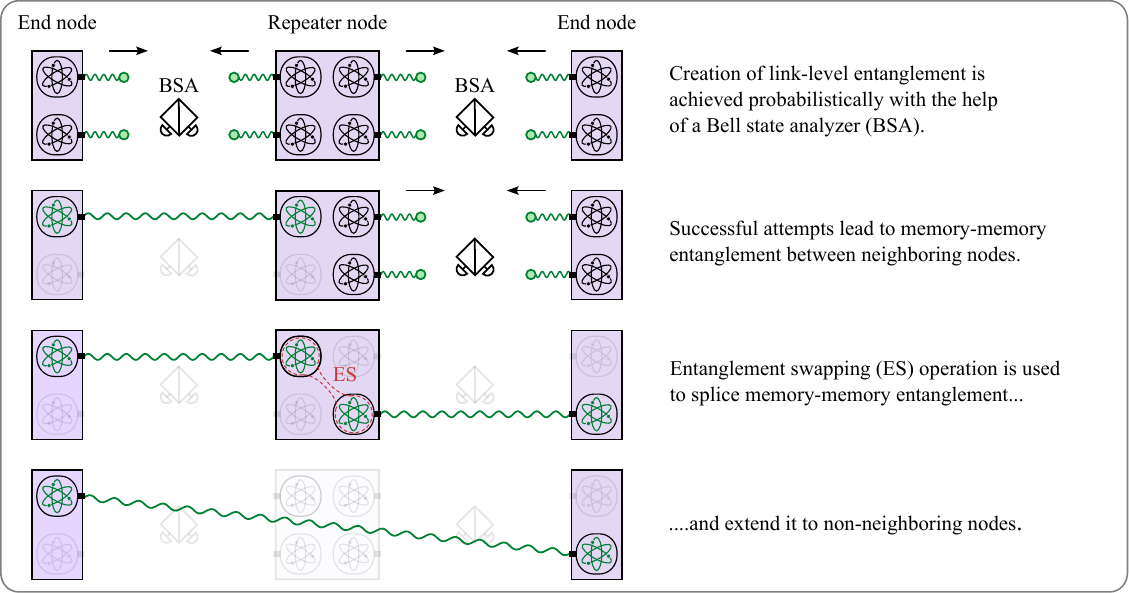}
    \caption{Basic principle behind memory-based quantum repeaters. Link-level entanglement is stored in quantum memories, represented by the atom symbol, until a neighboring link succeeds in its entanglement generation attempt. Entanglement swapping is used to extend this link-level entanglement to an end-to-end entanglement.}
    \label{fig:repeaters-overview}
\end{figure*}

The quantum repeater~\cite{briegel-repeater} stands as a cornerstone in the development of quantum communication networks, enabling the distribution of entanglement over long distances. 
At its core, the quantum repeater operates by splicing two shorter Bell pairs into a longer Bell pair through a process known as \textit{entanglement swapping}.
To establish an end-to-end Bell pair, the process begins with the generation of link-level entanglement -- a Bell pair between two adjacent nodes.
Through a series of entanglement-swapping iterations, this link-level entanglement is transformed into an end-to-end Bell pair, as shown in Fig.~\ref{fig:repeaters-overview}.

Quantum repeaters can be classified into three generations~\cite{azuma-rmp-repeater-review} based on how they manage quantum operational and photon loss errors (see Table~\ref{table:generations-of-repeater-table} for more details).
First-generation repeaters (1G) manage both errors in a heralded fashion.
Second-generation repeaters (2G) still manage the loss error via a heralded approach while using quantum error-correcting codes to manage quantum operational errors~\cite{jiang-2g-repeater}.
Third-generation repeaters (3G) circumvent both types of errors via quantum error correction.
Generating shared entangled states is accomplished via distributed computation in 1G and 2G networks.
Unique feature of 3G networks is that it is also possible to create store-and-forward quantum packet switching transmission of quantum data.
This is done by decoding and re-encoding the encoded photonic quantum states at every repeater station~\cite{borregaard-3g-repeater}.
2G networks place very high demands on link-level entanglement generation rates, and local gate and memory operation fidelities.
3G networks require high photon detection probability and local gate fidelities.
The all-photonic quantum repeater~\cite{azuma-rgs} is classified as 3G, placing rather different demands on the development of hardware components.

\begin{table*}[t]
\centering
\setlength{\extrarowheight}{2pt}
    \begin{tabulary}{\textwidth}{|c|L|L|L|}
        \hline
        \textbf{Generations} & \textbf{Operational error suppression scheme} & \textbf{Loss error suppression scheme} & \textbf{Explanation} \\
        \hline \hline
           1G repeater & 
           Probabilistic heralded entanglement purification – multiple copies of Bell pairs shared between the two parties can be consumed to increase the confidence that the remaining pairs are less likely to have errors (error detection). &
           \hfil
           \multirow{2}{=}{Probabilistic heralded link-level entanglement generation -- measurement devices (BSA) notify the two neighboring repeaters of the success-or-failure of the link generation.} & 
           No redundancy of any kind is used in the 1G repeaters\cite{briegel-repeater} , thus being the most efficient in terms of entanglement resources to the number of memories. Since errors are suppressed in a heralded fashion, detecting operation errors of non-neighbors will introduce communication delay leading to slower end-to-end rates. \hfil \\
        \cline{1-2}\cline{4-4}
           2G repeater &
           \multirow{2}{=}{Errors are detected and corrected via the redundancy added (either via multiple memories or multiple photons) through quantum error-correcting code.} &
           &
           Additional quantum memories may be used to encode multiple qubits into one Bell pair via quantum error-correcting code. In contrast to 1G repeaters, 2G repeaters do not require two-way communications between non-neighbor nodes as operation errors are determined locally at each repeater~\cite{jiang-2g-repeater}. If the operation errors do not accumulate over multiple hops, 2G may not require encoding and thus be 1G without error detection or correction.\\
        \cline{1-1}\cline{3-4}
           3G repeater &
           &
           Photons are encoded into logical qubits and sent through the optical fiber channel – provided that the code and the code distance are chosen appropriately the link-level generation is near-deterministic. &
           3G repeaters require the most resource-intensive to operate. In addition to possibly encoding multiple memories into one qubit, photons sent to measurement devices are also encoded~\cite{azuma-rgs}. This near deterministic suppression of both errors allows certain implementations of 3G repeaters to relay quantum states in packet-forwarding style~\cite{borregaard-3g-repeater}. \\
\hline
\end{tabulary}
\caption{Different generations of quantum repeaters have different ways of dealing with operational and loss errors.}
\label{table:generations-of-repeater-table}
\end{table*}

It should be noted that generations with higher numbers do not imply a better performance.
Each generation has its parameter regime of hardware characteristics where it is expected to perform best.
Thus, the generation of repeaters is a guide to how one would choose to build a network based on the equipment at hand.

The choice between all-photonic and memory-based repeater schemes depends on several considerations.
Memory-based repeaters are favored when high-quality quantum memories are available.
This means the memories have long coherence time, high-fidelity
\footnote{Fidelity is a measure of how close the actual physical state or operations are to the ideal ones.}
local gates, and can be coupled to optical fibers, as detailed later in challenges and open problems section.
Conversely, all-photonic schemes are better suited for scenarios where good quantum memories are unavailable but high-quality photon sources and adaptive measurement devices are.

\section{All-photonic quantum repeaters}

Interestingly, the entanglement swap does not need to be performed after link-level entanglement is created.
It can be performed first, provided that the probability of link-level generation is high enough.
This can be achieved by introducing redundancy in the number of trials encoded via a particular quantum state known as the \textit{repeater graph state} (RGS).
This ``time-reversed'' procedure is the core concept of the all-photonic quantum repeater~\cite{azuma-rgs}.
In subsequent discussions, we will refer to this repeater scheme as the \emph{RGS scheme}.

\subsection{Repeater graph state}

\begin{figure*}
    \centering
    \includegraphics[width=1\linewidth]{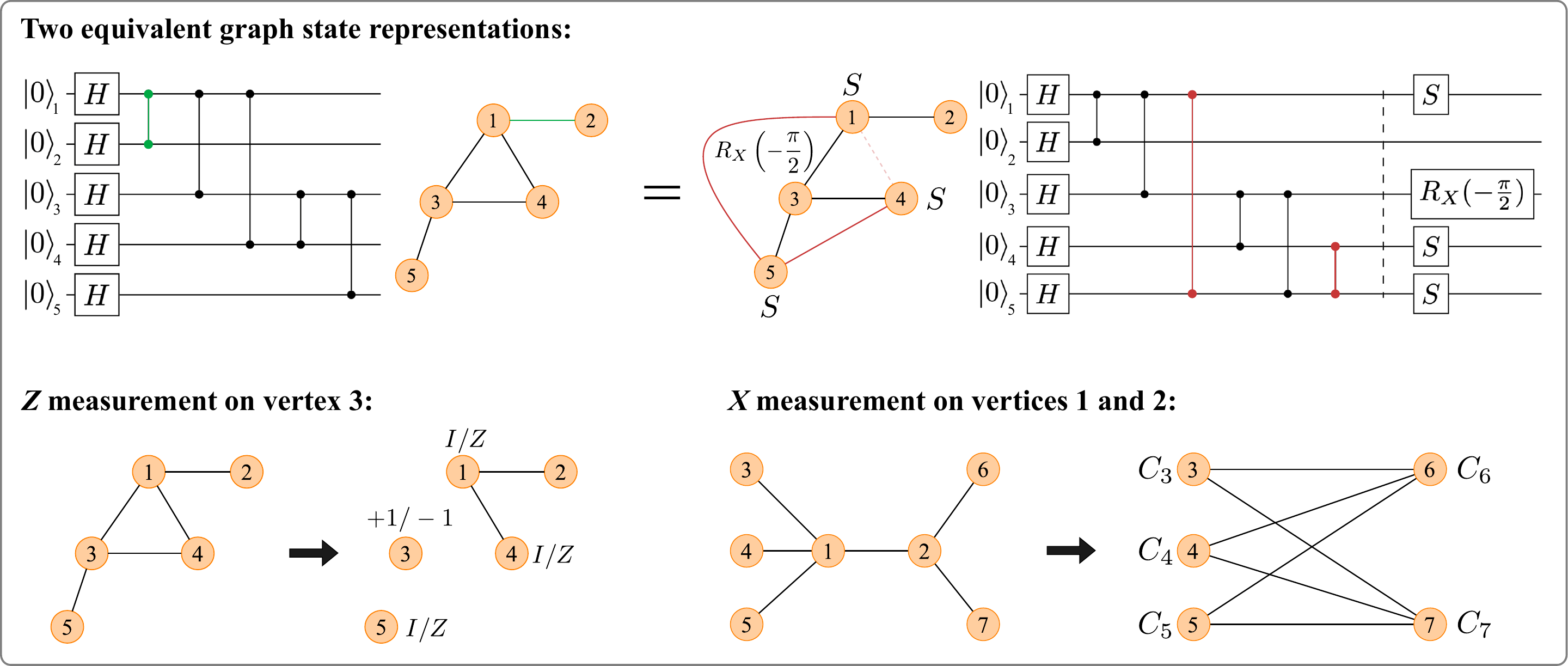}
    \caption{The same graph state can be described in multiple equivalent ways. Vertices of a graph correspond to qubits in the quantum circuit. Applying a controlled phase gate between two qubits corresponds to an edge between the vertices. One such example is shown in green in the top left. The resulting graph state can be represented by applying local complementation on vertex 3, as shown in the top right. This deletes any existing edges between the neighboring vertices of vertex 3, in this case edge $(1,4)$, and creates new ones that were previously missing, edges $(1,5)$ and $(4,5)$ in red. Applying the shown Clifford operations ensures that despite a different graph representation, the quantum states are the same in both cases. Visualization of a Z measurement is shown in the bottom left, while the effect of two X measurements (XX measurement) on a different graph is shown in the lower right. The Clifford operations $C_i$ are either $I$ or $Z$ depending on the outcomes of the two measurements.}
    \label{fig:graph-state-examples}
\end{figure*}

The workings of the RGS scheme are best explained via the language of graph states.
Graph states~\cite{hein-graph-state-pra} are a class of quantum states that have an intuitive description in terms of mathematical graphs, as shown in Fig.~\ref{fig:graph-state-examples}.
Each vertex in the graph corresponds to a qubit while the edges connecting them correspond to an entangling operation between the two qubits.
Graph states belong to a special class of quantum states that have an efficient description that scales quadratically in the number of qubits.
Manipulations of these graph states admit efficient and compact mapping as long as the operation maps one graph state to another.
In this picture, vertices that belong to the same connected component are parts of the same entangled system.
A Bell pair, viewed from the graph state formalism, is a two-vertex graph with an edge connecting the two vertices.

The graph state representation of a quantum state exhibits a non-uniqueness of its description.
Two different graphs can describe two equivalent graph states, provided the two states are related by application of local Clifford operations\footnote{Single qubit Clifford gates encompass rotations around the X, Y, or Z axis of a single qubit, with the rotation angle being an integer multiple of $\pi/2$.}. 
The Clifford operator acting on each vertex is called a \textit{vertex operation} (VOP), shown in Fig.~\ref{fig:graph-state-examples} as a letter next to the vertex except for when the VOP corresponds to the identity operator $I$.
The application of single qubit Clifford operators to a graph state can result in a distinct graph state representation, possibly changing its original edge structure and VOPs.
Notably, employing local complementation\footnote{Local complementation on a vertex $v$ of a graph is obtained by complementing or inverting a subgraph induced by the neighbors of $v$. Simply put, for every two neighbor vertices of $v$, delete the edge if they are previously connected or join them with an edge if they are not.} on a graph induces a graph with an alternative description of the same quantum state, as illustrated in Fig.~\ref{fig:graph-state-examples}.

Measuring a qubit in a Pauli basis results in another graph state. 
In this article, we only need to familiarize ourselves with Z basis measurement and X basis measurements performed on two adjacent vertices that share no neighbors, also called XX measurement. 
The effect of these two sequences of measurements is shown in Fig.~\ref{fig:graph-state-examples}.

The RGS comprises $2m$ physical qubits, referred to as \textit{outer qubits}, and $2m$ logical qubits, referred to as \textit{inner qubits}.
The inner qubits are arranged in a complete graph, with the outer qubits being connected to a single inner qubit, as shown in Step 1 of Fig.~\ref{fig:rgs-scheme-overview}.
RGS is defined by the parameter $m$ and a branching vector $\vec{b}$, where $m$ refers to the number of arms and $\vec{b} = (b_1, b_2, \ldots, b_{n})$ denotes the logical tree encoding of the inner qubits used for counterfactual measurement.
These two parameters determine how much loss and error the RGS can tolerate.

\subsection{RGS scheme}

\begin{figure*}[htb]
    \centering
    \includegraphics[width=\textwidth,keepaspectratio]{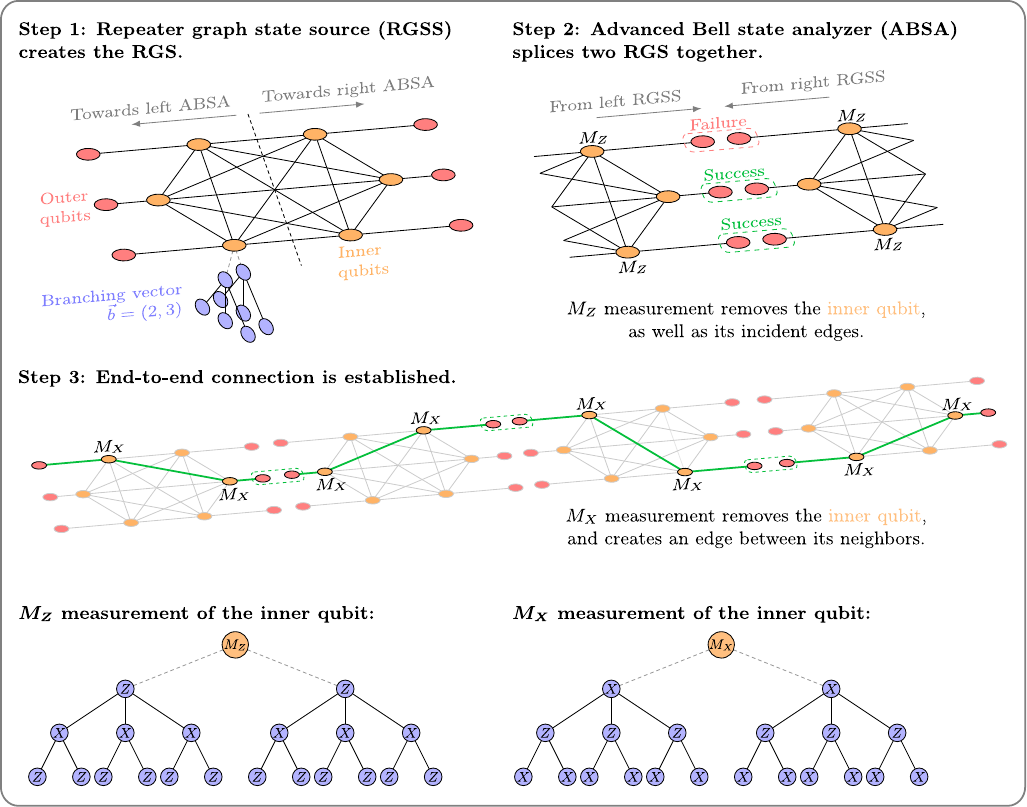}
    \caption{An overview of the RGS scheme. The three steps shown here have corresponding actions to the memory-based repeater scheme, where inner encoded qubits correspond to the memories while outer qubits correspond to the emitted photons. RGS generation in step 1 (at RGSS) mirrors the entanglement swapping of quantum memories but without actually choosing which inner qubits will be paired up. Step 2 (at ABSA) illustrates the link-level generation process through the BSM between each pair of outer qubits. The Z measurement on inner qubits in step 2 and the X measurements in step 3 signify the choosing of which pairs are swapped. Logical measurements of inner qubits are depicted at the bottom for both Z and X basis measurements. The $Z$ and $X$ labels inside the blue physical qubits indicate the actual physical measurements.}
    \label{fig:rgs-scheme-overview}
\end{figure*}

In the RGS scheme, the previously designated repeater nodes and BSA nodes are redefined as RGS source nodes (RGSS) and adaptive measurement nodes, or advanced Bell state analyzer nodes (ABSA).
As suggested by their names, RGSS generates the photonic RGS and transmits them to ABSA for measurement.

The RGS is generated at all RGSSs along the path between the two end nodes intending to share a Bell pair, as depicted in Step 1 of Fig.~\ref{fig:rgs-scheme-overview}.
At each RGSS, the RGS undergoes a split into two halves, with one half directed to the left ABSA and the other to the right ABSA.
Each ABSA receives the halves of two distinct RGSs and executes three stages of adaptive measurements.

In the initial stage, ABSA performs rotated Bell-state measurement (BSM) between outer qubits from the left and right RGS, as shown in Step 2 of Fig.~\ref{fig:rgs-scheme-overview}.
This rotated-BSM is akin to initially connecting the two vertices with an edge and subsequently performing the XX measurement on them.
This process is repeated for all $m$ pairs. 
Given the inherent probabilistic nature of the BSM\footnote{BSM implemented via linear optics has a theoretical limit of 50\% success probability without introducing ancilla photons.}, the success or failure of the outer qubits' BSM is uncertain, compounded by the possibility of photons being lost in the optical fiber during the journey from the RGSS to the ABSA.

If at least one pair of outer qubits is successfully measured in the first stage, ABSA moves on to the second stage.
Measuring a qubit in the Z basis is equivalent to removing the corresponding vertex from the graph state, as shown in Fig.~\ref{fig:graph-state-examples}.
Therefore, Z measurements are performed on the inner qubits connected to the outer qubits that failed the BSM, removing them from the RGS.
If all BSMs fail in any ABSA along the connection path, the entire connection attempt is deemed a failure and must be restarted.
The inner qubits linked to successfully measured outer qubits also undergo Z measurements, excluding a single pair.
Following the completion of all Z measurements, the resultant state forms a linear chain graph state.

Step 3 of Fig.~\ref{fig:rgs-scheme-overview} shows the concluding stage, where all inner qubits in the linear chain graph state undergo measurement in the X basis, constituting the XX measurement since only two qubits are left in each ABSA at this point.
The outcome is a locally equivalent Bell pair shared between the two end nodes.

\subsection{Logical measurements of inner qubits}

Logical encoding of the inner qubits is an essential part of the RGS scheme that allows us to forego quantum memories. 
The RGS scheme hinges on measurements of all inner qubits to be successful.

The tree encoding of inner qubits is mainly to combat photon loss.
The logical measurement operation on inner qubits is achieved by measurement of physical qubits, where the basis of measurements of all qubits in the same level of the tree is the same, and the basis of measurement changes depending on the even or odd level of the tree, as shown in the bottom panel of Fig.~\ref{fig:rgs-scheme-overview}.
To perform a logical Z measurement, qubits in the odd levels of the tree are measured in the Z basis while those in the even levels are measured in the X basis.
Similarly, for the logical X measurement, measurement bases are X in the odd levels and Z in the even levels.
The success of these logical X and Z measurements is determined via the success or failure of the first level of physical qubit measurements. 

In this tree structure, the Z measurement of the physical qubits inside the tree can be indirectly measured -- the measurement is considered successful even if the photon for this qubit is lost via a deduction of measurement results of its neighbors and their neighbors at one and two levels below it. 
It is also worth noting that the success probabilities of logical Z and X measurements are different. 
As seen from Fig.~\ref{fig:rgs-scheme-overview}, a logical Z measurement is successful if all the first level Z measurements on physical qubits, either direct or indirect, are successful.
A logical X measurement succeeds if at least one of the X measurements at the first level succeeds.
This leads to a nontrivial tradeoff between increasing the number of arms $m$ to increase the link-generation probability and the probability of successfully measuring all inner qubits.

\subsection{Advantages of the RGS scheme}

The RGS scheme offers a higher Bell pair generation rate compared to the traditional memory-based approaches.
This is due to the time-reversed protocol since repeaters no longer need to hold quantum memories awaiting messages heralding link-level generation before entanglement swapping can be performed.
The generation rate is only limited by the RGS generation time, thus a link shared between two RGSS nodes is only occupied by this generation time and can be reallocated for another connection much faster.
The waiting time at end nodes, on the other hand, is still limited by the same duration as memory-based repeaters: the time it takes for a message from the furthest ABSA to arrive.
It is not yet clear how shortening the busy time of the RGSS and ABSA nodes affects the overall performance and the service quality of a network.
This suggests that figures of merit for routing in the RGS scheme might be different from those of memory-based repeaters.

\section{Classical Communication for the RGS Scheme}

Current research efforts prioritize the optimal use of quantum resources~\cite{hilaire-rgs-optimizing-gen-time} (e.g., number of photons in the RGS, how the RGS is generated, or tradeoff between end-to-end Bell pair rate against the number of quantum emitters) while treating classical messages as a free resource.
In a real-world implementation, the network bandwidth that these classical messages need to take and their processing could become the bottleneck of the end-to-end Bell pair generation rate.

Consider end nodes separated by 1,000 km with RGSSs located every $\sim4$ km, and fiber attenuation of $0.2$ dB/km.
The RGS structure that maximizes the Bell pair rate per emitter is $m = 14$ and $\vec{b} = (10, 5)$, as shown in~\cite{hilaire-rgs-optimizing-gen-time}.
The total amount of classical information that needs processing is 545,462 bits.
Various quantum network applications require Bell pairs generated at megahertz rates, thus half a terabit of information needs to be processed per second to prepare these Bell pairs.
In the original proposal of all-photonic repeaters~\cite{azuma-rgs}, all this information is to be communicated to the end nodes for processing.
In this section, we propose to process this information in a distributed fashion in order to decrease the total load on the end nodes.

\subsection{Clifford Side Effects (VOPs) of the RGS}

Until now, we have assumed that the RGSSs are capable of generating the RGS without any VOPs, which is not always the case.
Generation of the RGS either via photonic fusion~\cite{azuma-rgs} or deterministic quantum emitters~\cite{buterakos-graph-generation, hilaire-rgs-optimizing-gen-time} introduces possible VOPs to most of the physical qubits since both rely on joining smaller graph states via measurements.
Therefore, one can either first apply operations to the photons at RGSS before sending them out to ABSA or track all the VOPs of all the qubits and send this information to be processed at end nodes.
We consider the latter approach, where the photons are sent to ABSA immediately after being generated and the information regarding the VOPs is sent later.

The crucial part is that the ABSA does not need to wait for this VOP information in order to decide which basis to measure in, provided that the possible set of VOPs is restricted.
This is true in the case of the RGS generation proposed in~\cite{naphan-rgs-protocol}, where the VOPs are restricted to be $I$ or $Z$.
ABSA performs Pauli Z and X measurements on the inner qubits.
Only the X measurement results get flipped if the qubit has $Z$ as its VOP\footnote{The Pauli Z operator ($\pi$ rotation around the Z axis) inverts the directions of the X axis.}.
The measurement outcomes are the same as having no VOPs in all other cases.
This implies that the ABSA does not need to know the VOPs of qubits in advance to select the basis of measurements and can follow the steps mentioned previously without any modification.

\subsection{Transmission order of photons in the RGS}

The basis selection of the inner qubits is dependent on the BSM outcome of the outer qubits.
It can be seen that as long as the outer qubits arrive at the ABSA before their connected inner qubits, the measurement basis can be determined locally by the ABSA.
The physical qubits composing the inner qubits can be sent in any order as long as it is known to the ABSA beforehand.
We note that even if the photons are lost in the fiber, assuming that the photons are well separated temporally, the ABSA can deterministically flag the loss event.
This well-separated assumption is also commonly adopted in memory-based repeater schemes for multiplexing photons from multiple memories into a single fiber~\cite{rdv-qi-architecture}.

\subsection{One-Stage versus Two-Stage Correction Method}

We now consider the number of classical bits that needs to be processed to obtain a single deterministic end-to-end Bell pair.
Each measured photon produces two bits of information; the measurement outcome, and the VOP.

\begin{figure}
    \centering
    \includegraphics[width=\columnwidth,keepaspectratio]{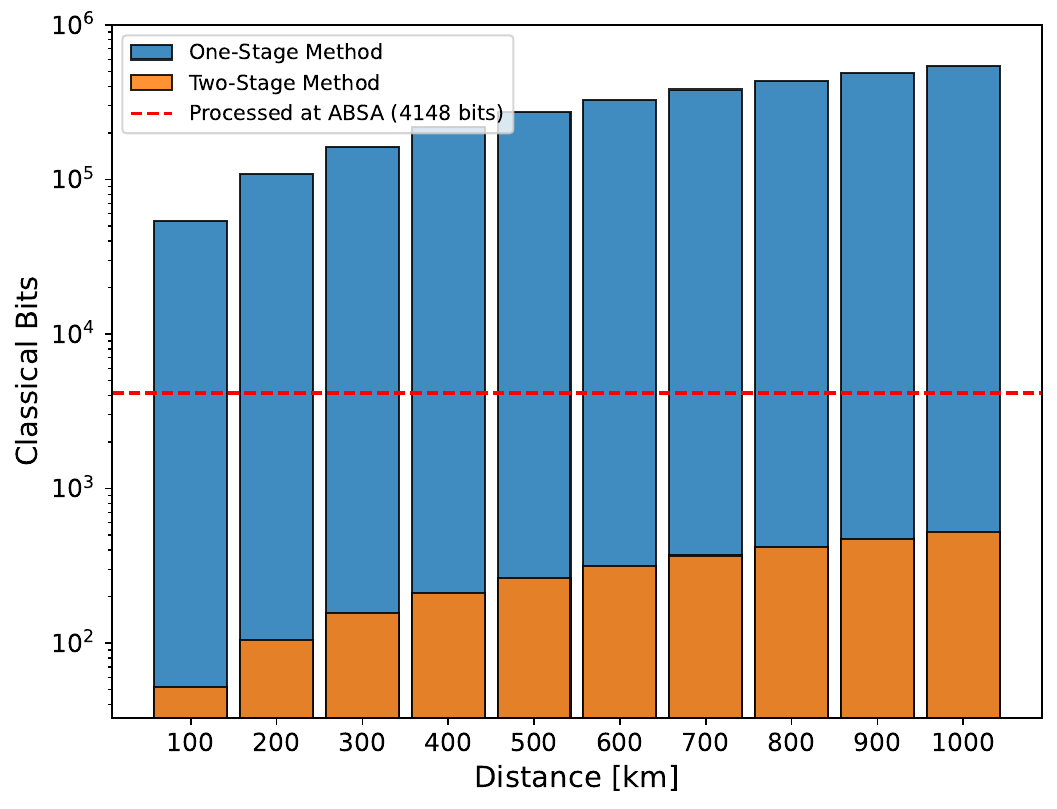}
    \caption{The number of classical bits to be processed, which is equal to the total number of photonic qubits to generate one Bell pair, with respect to the separation distance between two end nodes using the near-optimal RGS structure with $m = 14$ and $\vec{b} = (10, 5)$. Blue bars show the number of bits that the end nodes must receive and process to obtain the correction operation if all information is sent to end nodes only. Orange bars show the number of bits to be processed at the end nodes if the correction operations are calculated in a distributed, two-step process; in this scheme, each ABSA first processes the fixed amount shown by the dotted red line.
    }
    \label{fig:communication-cost}
\end{figure}

One method of correcting the VOPs is to communicate the classical information to the end nodes.
This is the usual method in literature~\cite{azuma-rgs}, and we refer to it as the \textit{One-Stage Correction Method}.
We consider an optimized structure of the RGS from~\cite{hilaire-rgs-optimizing-gen-time} in terms of the number of outer qubits $m$ and the encoding parameter $\vec{b}$.
This structure tries to maximize the Bell pair generation per quantum emitter given the distance between the end nodes.
The total number of classical bits per Bell pair that need to be communicated to and processed by the end nodes is shown in Fig.~\ref{fig:communication-cost} by the blue bars.

To lessen the communication and processing load placed on the end nodes, we propose a \textit{Two-Stage Correction Method}.
First, each ABSA gathers the VOP information from its adjacent RGSSs and reduces it down to only 2 bits of information; the two outcomes of X measurements on the appropriate inner qubits.
Next, these two bits are sent to be processed at the end nodes, calculating which correction operations are required to be performed to obtain the correct Bell pair.

Splitting the correction process into two stages has two clear advantages.
First, the bulk of the classical information generated by the RGS scheme does not need to be communicated to the end nodes, reducing the total load on the network and on the end nodes themselves.
Second, by processing the generated classical information in a distributed fashion at the ABSAs, the end nodes can compute the final correction faster and with less effort.
The new amount of classical information that has to be received and processed by the end nodes is represented by the orange bars in Fig.~\ref{fig:communication-cost}, showing an improvement of three orders of magnitude compared to the One-Stage Method.
The Two-Stage Method incurs only a constant processing load on the ABSAs, as shown by the red dashed line in Fig.~\ref{fig:communication-cost}.

\section{Challenges and open problems}

Although the all-photonic repeater is well understood at the theoretical level, there are engineering challenges in designing protocols between network nodes and other open problems, where further research is needed to improve the practicality of the scheme.

\emph{Generation of RGS. ---} 
Creating highly entangled states with flying photonic qubits is difficult.
One approach uses the \emph{photonic fusion operator}, which is a probabilistic process for entangling two multi-photon states by performing a joint measurement on two of their photons.
This approach imposes a large overhead penalty if the success probability of fusion is low.
Another way is to create the RGS deterministically via quantum emitters where entangling gates between emitters can also be deterministically applied~\cite{buterakos-graph-generation}.
In theory, this is more efficient and would enable a faster generation rate but the current hardware lacks the required functionality.
Recent experimental advances in photonic fusion gates~\cite{lodahl-entangle-photonic-fusion}, which can also be used to entangle quantum emitters, suggest that the two approaches do not need to be mutually exclusive.

\begin{table*}[t]
\centering
\setlength{\extrarowheight}{2pt}
    \begin{tabulary}{\textwidth}{|C|C|C|C|C|C|}
        \hline
        \textbf{Quantum memory/emitter} & \textbf{Coherence time} & \textbf{Gate fidelity} & \textbf{Coupling efficiency} & \textbf{Atom-photon fidelity} & \textbf{Emitted wavelength (nm)}\\
        \hline \hline
        Neutral atoms & 2.6 ms & $\geq 97.5\%$ & $60\%$ & $89\%$ & 780\\
        Trapped ions & $>$ 1 h & $\geq 99 \%$ &  & $86\%$ & 369 (\ce{^{171}Yb+}), 493 (\ce{^{128}Ba+})\\
        Quantum dots & \SI{3}{\micro s} & $95\%$ & $57\%$ & $\geq 80\%$ & 900--1565\\
        Defects in diamond (\ce{N}-V) & 0.6 s & $\geq 99\%$ & $37\%$ & $96\%$ & 637 \\
        Defects in diamond (\ce{Si}-V) & 10 ms &  & $85\%$ & $94\%$ & 737 \\
\hline
\end{tabulary}
\caption{Characteristics of state-of-the-art candidates for quantum memories and emitters used in deterministic RGS generation. Experimental data was taken from~\cite{azuma-rmp-repeater-review, hilaire-rgs-optimizing-gen-time}.
}
\label{table:memory-characteristics}
\end{table*}

Table~\ref{table:memory-characteristics} summarizes state-of-the-art experimental data on quantum memories~\cite{azuma-rmp-repeater-review, hilaire-rgs-optimizing-gen-time}, which play a pivotal role in both memory-based and RGS scheme implementations with deterministic generation approaches. As seen from the table, different memory types exhibit different characteristics. There is no clear winner between RGS scheme repeaters and memory-based repeaters, nor is there a clear preference among different generations of repeaters. However, the RGS scheme performs best when end-to-end coupling efficiencies exceed 85\%, gate fidelity approaches ideal, and emitter coherence times exceed 2500 times the gate duration~\cite{hilaire-rgs-optimizing-gen-time}.

\emph{Optimization of RGS structure depending on link characteristics and connection paths. --- }
In a complex network, the path between two nodes is not fixed and is determined upon request via a routing algorithm.  However, the RGSS decides the two parameters $m$ and $\vec{b}$ based on path characteristics.
Furthermore, if the error characteristics of each link are different, determining a good RGS structure is necessary to maintain good services of the network.
Currently, the optimal RGS structure for a given separation distance with RGSS and ABSA nodes evenly placed is found via exhaustive search~\cite{hilaire-rgs-optimizing-gen-time}.
There is a need for algorithms that find good RGS structures on the fly for each connection path and can also adapt to changes in link characteristics.

\emph{Time synchronization between nodes in the network. --- }
In order for the link-level generation to succeed, the two photons sent to ABSA for BSM (one from the left and another from the right) need to arrive at the same time\footnote{The two photons need to be indistinguishable for the BSM to be successful. Thus, this time window that can be considered as ``same time'' varies with the efficiency of photon detectors and also the properties of photons themselves.}. 
For the memory-based quantum repeaters, time synchronization is required only between each pair of repeaters that share a link because the generated Bell pairs are then stored in memories.
On the other hand, the time synchronization for the RGS scheme is more complex if the end-to-end RGS is created in one go, leading to the need to synchronize the clocks for all the RGSS nodes and ABSA nodes along the connection path.
This problem can be alleviated if certain parts of RGS generation are delayed, causing a drop in the generation rate, or the two halves of the RGS, which need to be sent to different ABSAs, are generated independently.

\emph{End node participation. ---} 
So far, research has focused mainly on the RGSS and the ABSA, making it unclear how end nodes should participate in the case where the generated Bell pairs are to be used for applications beyond quantum key distribution. 
These applications often require end nodes to have actual corrected Bell pairs. 
The Bell pairs need to be stored in quantum memories, as opposed to just directly measuring the RGS and processing only the classical outcomes for secret key generation.
One approach would require end nodes to have the same number of available quantum memories as the number of RGS arms, all emitting photons to the ABSA.
Upon success, only one of the memories holds the generated Bell pair, wasting the capacity of storing more entangled states in the memories.
A better approach would be to create just half of the RGS and anchor it to the memory.

\emph{Integration with memory-based repeaters. ---}
It is often thought that the all-photonic quantum repeaters and the memory-based repeaters are incompatible with each other.
The approach for including end nodes in the RGS scheme could bridge this incompatibility.
Due to the high generation rate and short busy time of RGSS and ABSA, combining the two architectures could lead to a better service in a quantum network.
The research area of combining different types of repeaters in a single network is still an open question to be explored.

\emph{Routing, multiplexing, and multipartite state distribution. ---}
Although the basic service of a quantum network is to distribute Bell pairs, distributing multipartite states directly from the network instead of creating them at the application layer can be much more efficient, especially in the early stages where quantum memories have short coherence times.
It is unclear whether this could be efficiently achieved for the RGS scheme or not since the behavior of the RGS scheme is akin to the circuit-switched network and the interplay between routing and multiplexing is not fully understood in the more flexible operations of memory-based repeaters.
This creates a need for efficient quantum network simulators to study this emergent behavior, which is likely different from the patterns seen in classical networks.

The open problems and challenges listed here are by no means exhaustive.
Some of the above open problems are being addressed~\cite{naphan-rgs-protocol}.

\section{Conclusion}

The all-photonic quantum repeater represents an alternative and sometimes overlooked approach towards the realization of quantum repeaters.
We have provided the operational intuition as a foundation for understanding the topic.
However, many engineering challenges still await solutions.
We proposed a distributed approach to dealing with the large volume of classical information generated per Bell pair, making all-photonic repeaters more amenable towards practical implementation.
Moreover, we have highlighted various open questions and research opportunities that beckon exploration, serving as stepping stones toward the ultimate realization of robust and scalable quantum networks.

\section{Acknowledgment}
This work is supported by JST [Moonshot R\&D][JPMJMS226C-104].
ChatGPT, a language model developed by OpenAI, was used in the development of this article for grammar checking, paragraph rewriting, and overall text improvement without generating novel content.

\bibliographystyle{IEEEtran.bst}
\bibliography{IEEEabrv, bibfile}

\begin{thebibliography}{10}
\providecommand{\url}[1]{#1}
\csname url@samestyle\endcsname
\providecommand{\newblock}{\relax}
\providecommand{\bibinfo}[2]{#2}
\providecommand{\BIBentrySTDinterwordspacing}{\spaceskip=0pt\relax}
\providecommand{\BIBentryALTinterwordstretchfactor}{4}
\providecommand{\BIBentryALTinterwordspacing}{\spaceskip=\fontdimen2\font plus
\BIBentryALTinterwordstretchfactor\fontdimen3\font minus \fontdimen4\font\relax}
\providecommand{\BIBforeignlanguage}[2]{{%
\expandafter\ifx\csname l@#1\endcsname\relax
\typeout{** WARNING: IEEEtran.bst: No hyphenation pattern has been}%
\typeout{** loaded for the language `#1'. Using the pattern for}%
\typeout{** the default language instead.}%
\else
\language=\csname l@#1\endcsname
\fi
#2}}
\providecommand{\BIBdecl}{\relax}
\BIBdecl

\bibitem{wehner-vision-road-ahead}
S.~Wehner, D.~Elkouss, and R.~Hanson, ``Quantum internet: {{A}} vision for the road ahead,'' \emph{Science}, vol. 362, no. 6412, p. eaam9288, Oct. 2018. \href{https://dx.doi.org/10.1126/science.aam9288}{doi:10.1126/science.aam9288}

\bibitem{rfc9340}
\BIBentryALTinterwordspacing
W.~Kozlowski \emph{et~al.}, ``{Architectural Principles for a Quantum Internet},'' RFC 9340, Mar. 2023. \href{https://dx.doi.org/10.17487/RFC9340}{doi:10.17487/RFC9340}
\BIBentrySTDinterwordspacing

\bibitem{rdv-qi-architecture}
R.~Van~Meter \emph{et~al.}, ``A {{Quantum Internet Architecture}},'' in \emph{2022 {{IEEE International Conference}} on {{Quantum Computing}} and {{Engineering}} ({{QCE}})}.\hskip 1em plus 0.5em minus 0.4em\relax Broomfield, CO, USA: IEEE, Sep. 2022, pp. 341--352. \href{https://dx.doi.org/10.1109/QCE53715.2022.00055}{doi:10.1109/QCE53715.2022.00055}

\bibitem{briegel-repeater}
H.-J. Briegel, W.~D{\"u}r, J.~I. Cirac, and P.~Zoller, ``Quantum {{Repeaters}}: {{The Role}} of {{Imperfect Local Operations}} in {{Quantum Communication}},'' \emph{Physical Review Letters}, vol.~81, no.~26, pp. 5932--5935, Dec. 1998. \href{https://dx.doi.org/10.1103/PhysRevLett.81.5932}{doi:10.1103/PhysRevLett.81.5932}

\bibitem{azuma-rgs}
K.~Azuma, K.~Tamaki, and H.-K. Lo, ``All-photonic quantum repeaters,'' \emph{Nature Communications}, vol.~6, no.~1, p. 6787, Apr. 2015. \href{https://dx.doi.org/10.1038/ncomms7787}{doi:10.1038/ncomms7787}

\bibitem{hasegawa-passive-absa-experiment}
Y.~Hasegawa \emph{et~al.}, ``Experimental time-reversed adaptive {{Bell}} measurement towards all-photonic quantum repeaters,'' \emph{Nature Communications}, vol.~10, no.~1, p. 378, Jan. 2019. \href{https://dx.doi.org/10.1038/s41467-018-08099-5}{doi:10.1038/s41467-018-08099-5}

\bibitem{jian-wei-pan-rgs-experiment}
Z.-D. Li \emph{et~al.}, ``Experimental quantum repeater without quantum memory,'' \emph{Nature Photonics}, vol.~13, no.~9, pp. 644--648, Sep. 2019. \href{https://dx.doi.org/10.1038/s41566-019-0468-5}{doi:10.1038/s41566-019-0468-5}

\bibitem{azuma-rmp-repeater-review}
K.~Azuma \emph{et~al.}, ``Quantum repeaters: {{From}} quantum networks to the quantum internet,'' \emph{Reviews of Modern Physics}, vol.~95, no.~4, p. 045006, Dec. 2023. \href{https://dx.doi.org/10.1103/RevModPhys.95.045006}{doi:10.1103/RevModPhys.95.045006}

\bibitem{jiang-2g-repeater}
L.~Jiang, J.~M. Taylor, K.~Nemoto, W.~J. Munro, R.~Van~Meter, and M.~D. Lukin, ``Quantum repeater with encoding,'' \emph{Physical Review A}, vol.~79, no.~3, p. 032325, Mar. 2009. \href{https://dx.doi.org/10.1103/PhysRevA.79.032325}{doi:10.1103/PhysRevA.79.032325}

\bibitem{borregaard-3g-repeater}
J.~Borregaard, H.~Pichler, T.~Schr{\"o}der, M.~D. Lukin, P.~Lodahl, and A.~S. S{\o}rensen, ``One-{{Way Quantum Repeater Based}} on {{Near-Deterministic Photon-Emitter Interfaces}},'' \emph{Physical Review X}, vol.~10, no.~2, p. 021071, Jun. 2020. \href{https://dx.doi.org/10.1103/PhysRevX.10.021071}{doi:10.1103/PhysRevX.10.021071}

\bibitem{hein-graph-state-pra}
M.~Hein, J.~Eisert, and H.~J. Briegel, ``Multiparty entanglement in graph states,'' \emph{Physical Review A}, vol.~69, no.~6, p. 062311, Jun. 2004. \href{https://dx.doi.org/10.1103/PhysRevA.69.062311}{doi:10.1103/PhysRevA.69.062311}

\bibitem{hilaire-rgs-optimizing-gen-time}
P.~Hilaire, E.~Barnes, and S.~E. Economou, ``Resource requirements for efficient quantum communication using all-photonic graph states generated from a few matter qubits,'' \emph{Quantum}, vol.~5, p. 397, Feb. 2021. \href{https://dx.doi.org/10.22331/q-2021-02-15-397}{doi:10.22331/q-2021-02-15-397}

\bibitem{buterakos-graph-generation}
D.~Buterakos, E.~Barnes, and S.~E. Economou, ``Deterministic {{Generation}} of {{All-Photonic Quantum Repeaters}} from {{Solid-State Emitters}},'' \emph{Physical Review X}, vol.~7, no.~4, p. 041023, Oct. 2017. \href{https://dx.doi.org/10.1103/PhysRevX.7.041023}{doi:10.1103/PhysRevX.7.041023}

\bibitem{naphan-rgs-protocol}
N.~Benchasattabuse, M.~Hajdu\v{s}ek, and R.~{Van Meter}, ``Architecture and protocols for all-photonic quantum repeaters,'' 2023. \href{https://dx.doi.org/10.48550/arXiv.2306.03748}{doi:10.48550/arXiv.2306.03748}

\bibitem{lodahl-entangle-photonic-fusion}
Y.~Meng \emph{et~al.}, ``Photonic fusion of entangled resource states from a quantum emitter,'' Dec. 2023. \href{https://dx.doi.org/10.48550/arXiv.2312.09070}{doi:10.48550/arXiv.2312.09070}

\end{thebibliography}

\end{document}